\documentclass[sigconf]{acmart}

\AtBeginDocument{%
  }

\setcopyright{none}
\acmDOI{}
\acmISBN{}
\acmConference[LOCO2026]{2nd International Workshop on Low Carbon Computing}{September 10--11, 2026}{Lancaster, UK}

\begin{document}

\title{Enabling Organisational Change Through Ground-Up Initiatives:\\ A Case Study from the STFC Scientific Computing Department}

\author{Jessica Huntley}
\authornote{Both authors contributed equally to this research.}
\affiliation{%
  \institution{Computational Mathematics Theme, Rutherford Appleton Laboratory}
  \city{Didcot}
  \country{England}
}
\orcid{0009-0008-7651-6149}
\email{jessica.huntley@stfc.ac.uk}

\author{David McDonagh}
\authornotemark[1]
\affiliation{%
  \institution{CCP4, Rutherford Appleton Laboratory}
  \city{Didcot}
  \country{England}}
\orcid{0000-0003-0025-138X}
\email{david.mcdonagh@stfc.ac.uk}

\begin{abstract}
  Transforming digital research infrastructure (DRI) to align with UK Net Zero targets requires significant action from organisations in this space. Although high level strategies and recommendations exist, it is not always obvious how to translate these into concrete results. Here we present a case study from the Science and Technology Facilities Council's Scientific Computing Department (SCD). This department consists of over 200 staff supporting tens of thousands of researchers, and is spread over significant cloud and high performance computing infrastructure, as well as a diverse ecosystem of software across computational biology, materials science, engineering, and mathematics. We show how we developed the sustainability strategy for SCD across themes of emissions monitoring, user education, best practice, setting sustainability standards, and providing long-term support to sustainability work. We discuss the successes and difficulties in establishing this strategy, and how this could serve as a template for other departments and facilities.
\end{abstract}

\maketitle

\section{Introduction}
In 2021 the UK published its strategy for Net Zero, aiming to significantly decarbonise all sectors of the economy by 2050\cite{UKNetZeroStrategy}. This is reflected in the strategy of UK Research and Innovation (UKRI)\cite{UKRINetZeroStrategy}, and the strategy of the Science and Technology Facilities Council (STFC)\cite{STFCNetZeroStrategy}, of which SCD is a part. Specifically, the STFC Environmental Sustainability Strategy 2025-2030 aims to reduce overall emissions by embedding sustainability across broad themes of  governance, measurement and monitoring, culture and behavioural change, and leading by example\cite{STFCNetZeroStrategy}. This is badly needed in DRI. For 2024, it was estimated that electricity consumption from data centres was 1.5$\%$ of global demand, with trends driven by AI technology adoption meaning this could rise by approximately 15$\%$ per year from 2024 to 2030\cite{EnergyAI}. This is more than four times faster than the growth of total electricity consumption in all other sectors. In 2023, the UKRI DRI Net Zero Scoping Project estimated that UKRI's large scale computing facilities emit 40 kiloton CO$_2$e/yr\cite{DRINetZeroComputing}, the equivalent of flying 21 times around the world per day. The final report for the scoping project provided more than 200 recommendations for aligning DRI operations with Net Zero targets\cite{DRINetZeroScopingReport}. However, a strategy is needed to translate these recommendations to tangible actions for developers, facilitators, and users of DRI. 

A key part of this is the Network for sustainable Digital Research Infrastructure Vision and Expertise (NetDRIVE), which is bringing together creative thinking from across UKRI to build a common vision for a sustainable future, and bring about the transformative change required to embed sustainable working practices across DRI communities\cite{netdrive}. Below these cross-cutting principles, departments must translate these working practices into concrete actions in order for ambitious Net Zero targets to be reached. SCD is an interesting case study for this, as it maintains significant computing infrastructure and software, in addition to carrying out research analogous to a university. In 2022 the SCD Energy Efficiency Team (EET) was established to begin implementing broad recommendations in DRI across SCD services and software. Although initially a voluntary group, this is now partially funded by the department to design and carry out its sustainability strategy. Here we present a case study outlining how the EET is transforming SCD towards a computing department aligned with the challenges of Net Zero, and how these efforts fit in within the wider context of NetDRIVE and broader strategies. 

\section{Developing a Sustainability Strategy}

The EET initially operated as an informal monthly forum for interested people across SCD to discuss sustainable computing ideas, opportunities, and emerging best practice. At this time, initiatives were largely driven by individual enthusiasm and often limited in scope due to lack of dedicated funding. It soon became apparent, that in order to help drive progress towards ambitious Net Zero targets, a more structured approach was needed. The team spent time developing a sustainability strategy for the department, in order to formalise its objectives, and advocate for funding.

When designing the SCD sustainability strategy our focus was to use the broad principles in the UKRI\cite{UKRINetZeroStrategy} and STFC\cite{STFCNetZeroStrategy} strategies and adapt them based on the current state of our department.  This meant creating a record of all current and past sustainability work, and then identifying where we could have impact from the ground up, and what areas were out of scope. Ideally, the team would have first developed a detailed picture of the department's carbon footprint, and then prioritised initiatives according to their potential impact. However, obtaining accurate emissions data across the mix of compute services that the department maintains proved more complex than anticipated. Instead, the decision was made to pursue sustainability initiatives in parallel with developing tools to monitor and visualise emissions data rather than delaying action.

The five themes described in Section~\ref{strat} were chosen as they represented areas where the department could have a direct influence, and complement the work of the central STFC sustainability team. The STFC theme of encouraging culture and behavioural change,  for example, could readily be translated into setting up a framework for groups and themes in the department to apply for Green DiSC accreditation\cite{GreenDiSC}. Similarly, to address the theme of measurement and monitoring we argued for regular funding for graduate projects that we would supervise. Aspects that we considered out of scope included supply chains and wider organisational policies that were out of the control of the department.

As an alternative approach, the EET could have chosen to focus solely on technical projects, however considering the variety of work and areas of expertise across the department, the team felt better placed to function as a supporting layer. Here the idea would be to support teams in pursuing sustainability projects, provide long-term ownership of projects, and help foster collaboration.

\section{Strategy}\label{strat}

The SCD sustainability strategy designed by the EET is spread across five themes that broadly reflect the goals of the STFC Environmental Sustainability Strategy 2025-2030\cite{STFCNetZeroStrategy}.

\begin{figure}[h]
  \centering
 \includegraphics[width=\linewidth]{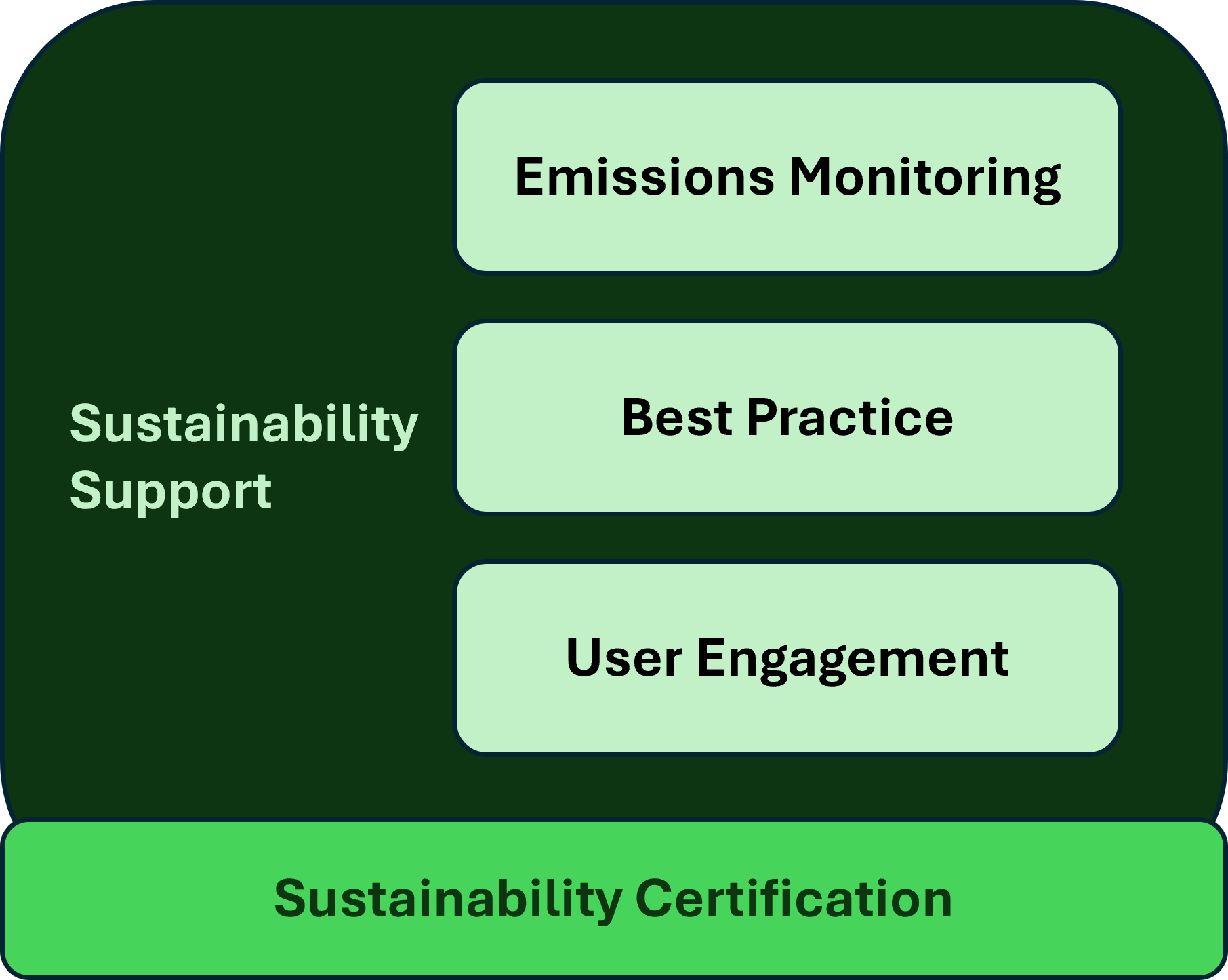}
  \caption{An illustration of the five projects that make up the SCD Sustainability Strategy.}
  \label{strategy map}
\end{figure}

As shown in Figure \ref{strategy map}, sustainability support acts as a coordinating layer for emissions monitoring, best practice, and user engagement, with each of these enabling sustainability certification. In the remainder of this section, we describe each of these themes in more depth, as well as current progress towards maximising energy efficiency.

\subsection{Sustainability Support}

Across large groups and departments, environmental sustainability initiatives are often uncoordinated, and this can lead to duplication of effort. Within SCD, it is also common for sustainability projects to be carried out by graduates or apprentices as part of their rotations. As they move on to other work, long-term responsibility for these projects is often missing. To address this challenge in SCD, the EET took on the role of a support hub for sustainability. Our aim is to connect related projects, foster collaboration, and maximise the impact of the work being carried out. 

To increase engagement across the wider department and maintain a level of awareness of current and upcoming sustainability-related projects, a network of sustainability representatives was established. SCD is an international centre of excellence for advanced computing expertise and digital research infrastructure, with skills and expertise spanning 12 themes, including computational science, infrastructure, data engineering, and software engineering. Each of the themes in SCD were asked to nominate a sustainability representative, to help disseminate information from the EET as well as provide updates on projects and initiatives within their theme. Initial engagement was good, with the EET successfully recruiting a representative from each theme. However, sustaining engagement has been challenging; often requests for information will receive limited responses. To address this, we are looking to introduce quarterly meetings where representatives are invited to share updates, improving accountability and sense of community. 

Without a clear map of existing activity around sustainability, it is difficult to identify opportunities for future work or collaboration. Through input from the sustainability representatives, the EET created a record of sustainability related projects across the department. So far this record has been used to identify projects to highlight through a ‘Sustainability Spotlight Series’, which is a collection of articles shared with the department through an internal SharePoint site. Projects have also been promoted externally through conference talks and posters, most recently at the 2025 CoSeC Conference \cite{CoSecEET}, NetDRIVE’s Oxford community meeting\cite{NetDriveCommunityMeeting}, and SC4RC 2026\cite{SC4RC2026}. 

In addition to collating information, it’s also important to create opportunities for people to share knowledge and discuss ideas. In September 2025, the EET organised a workshop to discuss sustainable computing across SCD, which featured lightning talks from 6 themes and 3 programmes, as well as updates from the EET and SCD senior leadership. The workshop highlighted the breadth of ongoing work across the department, and prompted cross-theme interactions. It was noted that whilst engagement was strong during the event, work must be done to sustain momentum. Making the workshop an annual or biannual event would help to address this. 

\subsection{Emissions Monitoring Across SCD Services}
In order to quantify the impact of sustainable computing initiatives across SCD, an understanding of the current carbon footprint associated with computing in the department is required. Whilst staff are generally interested in sustainable computing, often a lack of available time and funding to work in this area prevents progress from being made. To allow for dedicated time to embed emissions monitoring across SCD services, the EET have proposed and supervised several early career projects. SCD runs both apprentice and graduate schemes, which involve completing short-term (usually 6-month) projects across different themes in the department. The EET have found that organising early career projects has been an effective way to make significant progress in a short time frame, with the added benefit of training early career staff in sustainable computing skills.

As of June 2026, projects have focused on SCD HPC services, such as the SCARF and JASMIN clusters, and cloud services, such as STFC Cloud, where users typically use virtual machines (VMs).

SCD Apprentice George Roe developed the Green Usage Impact Logging Tool (GUILT), designed to monitor user emissions of HPC jobs on services like SCARF and ARCHER2\cite{GUILT}. GUILT provides a simple wrapper for any service using a SLURM queue\cite{slurm}, allowing users to track emissions of current and past jobs. It can also suggest to users the best time to run a job by looking at the carbon intensity forecast. This is now being used as part of another project, where SCD graduate James Ravindran is estimating the potential impacts of embedding green scheduling into SCD HPC services.

SCD Graduate Hieu Le worked with the STFC Cloud team to estimate the power usage associated with VMs using different models. He then designed dashboards allowing users to estimate carbon emissions associated with work done in STFC Cloud. This work has been built upon by SCD apprentice Ashraf Hussain, who has worked on displaying carbon emissions to users on SCD platforms that sit on top of STFC Cloud, such as Ada. The aim is now to bring together these projects under emissions profiles that allow users, groups, and themes to see their total emissions from SCD services. This is now being built upon graduate Thomas Croucher, with a project designed to dynamically visualise the sources of SCD emissions at different levels of granularity, and what impact different reduction strategies would have.

\subsection{Sustainability Certification}

Green DiSC \cite{GreenDiSC} is an open-access digital sustainability certification scheme, aimed at research computing groups, central teams, and research computing infrastructure teams. The scheme provides a roadmap for teams to tackle the environmental impacts of their work through a set of criteria co-developed with different stakeholders in DRI. As a tool for embedding best practice across the department, the EET is encouraging all themes and groups in the department to pursue Green DiSC certification. 

As of June 2026, there are three themes/groups in SCD that have achieved Green DiSC Bronze, including the EET, which is certified as a ‘central team’. The EET are now focused on using the resources developed through these successful applications to produce templates and supporting guidance for other interested groups. These documents have been made available to the rest of the department to reduce the effort required to submit an application. The EET also run webinars to introduce Green DiSC, outline the criteria and highlight support available. To encourage continual engagement with applications, the EET maintains a Jira board showing the status of different groups and themes, providing help where needed. Currently, support is focused on the Bronze criteria, but as Silver and Gold criteria are published, and the EET makes progress towards them, the resources and guidance offered to the rest of the department will be expanded. 

\subsection{User Engagement and Training}

The computing platforms and scientific software developed within SCD are used by several thousand users across the National Laboratories and the wider research community. The way users interact with these services can significantly impact their environmental footprint, and so it is vital to embed environmental sustainability considerations within user workflows. This is challenging in the current research climate, where sustainability does not often factor into decision making, commonly due to lack of time, incentives, or clarity on the best actions to take.

There is existing work that has set out broad principles of sustainable computing and proposed high-level recommendations \cite{DRINetZeroScopingReport, LannelongueGREENER}. However, focused training on reducing the environmental impacts of computing would be beneficial, providing targeted, practical information to facilitate change. The SCD EET is running the NetDRIVE DRI-EMIT (Digital Research Infrastructure Emissions Measurement and Interpretation Training) project in collaboration with STFC's Hartree Centre. This project aims to develop training modules that are tailored to different personas across DRI, including developers, service managers, facilitators, and users of compute services.

In addition to training, SCD Apprentice Ashraf Hussain explored how to communicate carbon monitoring information to users of the Ada platform as part of his project. The platform provides flexible VM access for scientists, with a comprehensive suite of pre-installed scientific software. Ashraf met with users of the service, determining how they interact with the service, and what information would help drive behaviour change. Following these conversations, proposed features include displaying 48-hour emissions forecasts and real-time usage graphs showing busy/idle time. This information would allow users to make data-driven decisions around when to schedule compute tasks and when to delete workspaces to reduce idle usage. The proof of concept design has been handed over to the Ada team, and another SCD apprentice, Jack Santiago, is now working on implementing the suggested changes.

\subsection{Best Practice and Knowledge Sharing}

Whereas user engagement focuses on enabling individuals to make informed decisions whilst interacting with existing systems, this relies on a clear understanding of what 'best practice' looks like. Part of the work of the EET involves keeping up-to-date with sustainable computing best practices, and signposting to sustainable computing tools and resources. 

Across SCD, the EET have initially focused on determining where expertise already exists, translating knowledge into easily digestible formats (such as wiki pages and cheat sheets) and connecting people and projects who could learn from each other. In March 2026, the EET conducted a survey across group leaders and service managers throughout the department in order to assess existing understanding of sustainable computing principles and identify commonly used technologies. An immediate action following this survey has been to reach out to those who have deemed themselves an expert in one or more sustainable computing concepts, to ask either for practical advice on the topic or a case study on how they apply it to their work. Where applicable, the team also plans to contribute content to open-access forums and repositories.

\section{Discussion and Conclusion}
\subsection{Outcomes and lessons learned}
The work in Section 3 outlines a strategy for reducing the environmental impacts of computing within SCD, and is designed to operate alongside wider organisational policy. Although high-level policy from the UKRI and STFC provide direction, impact is dependent on how policies are implemented across individual departments, groups, and services. The EET has aimed to translate broad principles into practical actions relevant to the work carried out in SCD.

Working to embed sustainability considerations directly into the services developed and managed by the department has been a particularly effective element of the strategy. We are now starting to offer carbon monitoring and usage feedback across SCD platforms, providing users with the information they need to understand and reduce their environmental impacts. Although there is some evidence to suggest that making users aware of their emissions can be enough to influence behaviour\cite{UserEmissionsStudy}, future work will look at different strategies to effectively reduce emissions at the user and group levels.

Another success of the strategy has been providing dedicated effort to centrally coordinate sustainability projects across the department. This has allowed us to improve visibility of work, and connect people and projects to avoid duplication of effort. Providing templates and resources for Green DiSC certification has also helped to lower the perceived effort of achieving certification and encouraged more teams to explore the scheme.

A key challenge in executing the strategy has been sustaining momentum and engagement with all themes in SCD. The EET started out as a working group with no dedicated funding, and although the team do now receive funding from both the department and NetDRIVE (for the DRI-EMIT project), overall funding and staff time are still very limited. Some elements of the strategy, for instance the sustainability representative scheme, could benefit from more structure and facilitation. There are also areas of sustainable computing that we have not yet considered as part of the strategy, for example sustainable procurement and embodied carbon of hardware, and monitoring the environmental cost of storing large amounts of data.

\subsection{Transferability}
Although this work is grounded in the context of SCD, many of the underlying principles are transferrable to other research computing teams and departments.

Existing frameworks like the UKRI Net Zero DRI roadmap \cite{DRINetZeroScopingReport}, GREENER principles \cite{LannelongueGREENER}, and Green DiSC \cite{GreenDiSC} provide guidance on environmentally sustainable computing practices. Green DiSC in particular provides complementary criteria for central (sustainability) teams, research groups, and research computing infrastructure teams. Three levels of certification (Bronze, Silver, and Gold) will be available and teams are expected to apply for recertification after 2 years, helping to sustain long-term progress. Whilst Green DiSC encourages working collaboratively on applications within an organisation, local coordination, advocacy, and support to embed sustainable computing principles within teams is still required.

Within SCD, the EET provides the necessary supporting function by connecting sustainability projects, maintaining visibility of existing work and developing shared resources, in order to reduce duplication of effort. The transferable aspect of the SCD strategy is therefore not the development of a new set of guidelines or framework, but the embedding of existing schemes and best practice standards across the work carried out by the department. The following activities provide a suggested structure for those looking to make change from the ground up, and could be applied irrespective of discipline:

\begin{itemize}
    \item Establish clear ownership of sustainability activities at a team, department, or organisational level through named sustainability representatives or a dedicated coordinating group. This should include maintaining a record of projects and sharing outcomes across internal (and if possible external) networks, where long-term responsibility for maximising the impact of this work is explicit.
    \item Provide standard tools and methodologies for measuring and reporting environmental impact, enabling consistency across the department/organisation. This could involve the development of specialised tools (e.g. the VM carbon calculator for the STFC Cloud) or guidance on using existing tools and frameworks.
    \item Create opportunities for knowledge sharing and community building. This might range from having informal discussions with colleagues or networking at conferences and events, to organising workshops and seminars or setting up a community of practice.
    \item Encourage the uptake of recognised standards and certification schemes. This could include dissemination of relevant schemes for your team or department, as well as the development of shared resources and guidance on local interpretation of scheme criteria.
\end{itemize}

The implementation of these recommendations can be adapted based on an organisation's size, structure, and available resources. Regardless of the exact approach taken, our experience within the EET suggests that the underlying principles of the strategy provide a foundation for driving positive change. These activities also provide practical actions which can help progress towards high-level organisational sustainability objectives.

 \section{Acknowledgments}

 This work has been made possible by funding from the STFC Scientific Computing Department, which funds 10\% of our time explicitly for sustainability projects. Training and best practice components are partly funded by NetDRIVE via the NetDRIVE DRI-EMIT project, in collaboration with the Hartree Centre. We are grateful to George Roe, Ashraf Hussain, Hieu Le, James Ravindran, Jack Santiago, and Thomas Croucher, all of whom have been vital in developing tools for monitoring SCD services. The sustainability strategy was written in collaboration with additional volunteers in the EET, including Elliott Kasoar, Harry Swift, and Anjali Bhatt. Finally, we are thankful to the teams running STFC Cloud, SCARF, and JASMIN, who have been supportive in supervising graduates and enabling projects, and for the leadership team in SCD who continue to fund and see value in this work.

\bibliographystyle{ACM-Reference-Format}
\bibliography{references}

\end{document}